\documentstyle[preprint,tighten,aps]{revtex}
\begin{document}
\preprint{INFNCA-TH-94-2}
\vskip1.5truecm
\draft
\title{Mass corrections in $\protect\bbox{J/\psi \to B\bar{B}}$ decay \\
and the role of distribution amplitudes}

\author{ Francesco Murgia}

\address{Istituto Nazionale di Fisica Nucleare,
Sezione di Cagliari \\
via Ada Negri 18, I--09127 Cagliari, Italy}

\author{Maurizio Melis}

\address{Dipartimento di Scienze Fisiche, Universit\`{a}
di Cagliari \\
via Ospedale 72, I--09124 Cagliari, Italy}

\date{December, 1994}

\maketitle

\begin{abstract}
We consider mass correction effects on the polar angular
distribution of a baryon--antibaryon pair created
in the chain decay process $e^-e^+ \to J/\psi \to B\bar B$,
generalizing a previous analysis of Carimalo.
We show the relevance of the features of the baryon distribution
amplitudes and estimate the electromagnetic
corrections to the QCD results.
\end{abstract}

\pacs{13.65.+i, 13.25.Gv, 12.38.Bx, 14.40.Gx}

\narrowtext

\section{Introduction}
\label{intro}

In the last few years our understanding of exclusive hadronic
processes at high transfer momentum,
in the framework of perturbative QCD, has improved
(for a comprehensive review and further references see, {\it e.g.},
\cite{brod89} ).
Theoretical models, based essentially on factorization ideas,
have been elaborated and refined.
The main ingredients of these models (which from now on
we shall indicate
as PQCD models) can be summarized as follows:
The amplitude for a given exclusive process is obtained
by convoluting two well distinct contributions:
the first coming from the
hard scattering among the partonic constituents
of the involved hadrons and the second from the
subsequent, soft processes which
lead to hadronization.
The hard scattering can be described by means of perturbative
QCD techniques, representing, in first approximation,
each participating hadron by
its valence constituents, assumed collinear with the parent hadron
and among themselves.
As for the soft processes, their treatment is outside
the possibilities of perturbative methods and alternative
approaches (like QCD--sum rules methods
or Lattice calculations) are required; in practice, they appear
in the so called hadronic distribution
amplitudes ($DA$) which describe, for each hadron (and independently
of the particular process under consideration), how its
momentum is shared among the valence constituents.
Although a complete, formal proof of the validity of these factorization
procedures is still lacking, at least for exclusive processes,
there is enough theoretical work which supports these models
\cite{brod89,coll87,coll89}.

As it should be clear, PQCD models acquire full validity
only at very high transfer momentum;
however, it is not yet clear what that means in practice
(that is, at what $Q^2$ scale we expect these
models to become reliable). This is a controversial
point \cite{isgu89}, even if very recently new developments seem to
justify the applicability of the models also for not so high
momenta \cite{ster92}.
On the other hand, we must not forget that all the cross sections
for exclusive processes behave, at high $Q^2$, as an inverse power
of $Q^2$. This power increases with the number of valence
constituents involved in the process and makes more difficult, from an
experimental point of view, to distinguish and measure
these increasingly rare events from the bulk of the inclusive
processes.
We may then summarize the situation as follows:
on one hand, theoretical models are surely
under better control at very high $Q^2$ but, on the other hand,
almost all the experimental information presently at our disposal
falls in a range of $Q^2$ which, while not completely
out of reach of PQCD techniques, it is not fully recognized
as an ideal laboratory for perturbative models.
In particular, it is not clear the role played by
higher order corrections (which are reflected in several possible
modifications of the basic PQCD models).
Unfortunately, the implementation of these higher twist
contributions is quite intricate.
It follows from what we said that a complementary,
theoretical and phenomenological analysis of
all the presently available experimental measurements
would be of great help in clarifying and
improving  PQCD models. Thus it is very
useful and important to undertake all the possible
efforts in order to shed light on controversial
points and improve our understanding of exclusive
processes.

We should also bear in mind that presently only for a few, relatively simple
processes, calculations have been performed. This is due to
the increasing complexity of calculations when more and
more hadrons (and, as a consequence, partonic constituents)
are involved in the process.
When compared with the experimental results
(all of which, with the possible
exception of the proton form factor,
are at intermediate values of $Q^2$), these calculations
show several successes but also some failures.
Most of these failures can be
attributed to violations of the so called helicity selection
rules \cite{brod89} (which are a specific property of PQCD models,
valid to all orders in the strong coupling constant
perturbative expansion)
and their overcoming requires the
introduction of higher twist effects.

Several attempts have been made in order
to implement the original PQCD models taking into
account higher order contributions, like higher valence Fock
states \cite{bena91}, transverse momentum effects \cite{ster92},
$L\neq 0$
angular momentum components in hadron wave functions
\cite{pire93},
constituent quarks mass effects \cite{anse90,anse92,anse93b},
diquark correlations inside baryons \cite{anse93c}.

For constituent quark mass effects, in particular,
a number of calculations \cite{anse90,anse92,anse93b}
have been performed for several experimentally observed
processes, some of which are allowed in PQCD models while others
are forbidden by the helicity selection rules.
In these calculations the elementary hadron
constituents are given a mass, $m_i = x_im_H$, where $m_H$ is
the hadron mass and $x_i$ is the (light-cone) fraction of the hadron
momentum carried by the $i$-th constituent,
opportunely weighted (in the convolution integral)
by the corresponding hadron distribution
amplitude. This effective mass could take into
account (in a global way) several higher order effects
which have been neglected in the ordinary PQCD models.

Although from a formal point of view this approach
requires further justifications, it offers a relatively
simple, parameter free, means for practical calculations.
The results obtained this way can be
compared with the lowest order PQCD results,
and, as we shall also see in the following, lead
to predictions for several effects that can be
experimentally tested at the present time or in the near future
(see also ref.~\cite{anse93b}).

Few years ago Carimalo \cite{cari87} considered
mass corrections effects on the polar angular
distribution of baryon--antibaryon pairs produced
in the exclusive decay of the $J/\psi$.
As it can easily be seen, PQCD models
predict, due to the helicity selection rules,
a distribution of the type $1 + \cos^2\theta_{_B}$
\cite{brod89},
where $\theta_{_B}$ is the polar angle which specifies the
direction of motion of the produced baryon in the
$J/\psi$ rest--frame. Although experimental results
are available only for a few baryons \cite{mark84,dm287} and are
in some cases affected by large statistical errors, there
are clear indications that the angular distributions
behave rather as $1 + a_{_B}\cos^2\theta_{_B}$,
with, {\it e.g.}, $a_p = 0.62 \pm 0.11$,
$a_\Lambda = 0.62 \pm 0.22$ \cite{dm287}.
Here $a_{_B}$ ($\leq 1$) is a factor which can
be expressed from the helicity amplitudes for the decay
process, as we shall see in detail in the next Section.

As shown in ref.~\cite{cari87}, mass corrections
can in principle explain why $a_{_B} < 1$ and lead to
a better agreement
between theoretical predictions  and experimental results.
Furthermore, since the parameter $a_{_B}$ is given as a ratio
of squared helicity amplitudes,
it is independent of the exact value of the baryon decay
constant, and of
several ``fine tuning'' details of the models
(for example, how to treat the strong coupling constant
in the convolution integrals); as such, their effects
may be neglected almost completely, as will be
clarified by our explicit calculations.

The results of ref.~\cite{cari87} are
obtained using a non--relativistic
bound--state model both for the decaying $J/\psi$ and the
produced baryons. However, while this approximation
is well grounded for heavy quark bound states, like the $J/\psi$,
it can be questionable for light hadrons.
In ref.~\cite{cari87} the proton decay constant was consistently
fixed in order to reproduce the $J/\psi \to p\bar p$
decay width.
A different, widely used approach (see {\it e.g.}
ref.~\cite{brod89,cern89} and references therein)
is to tentatively fix once and for all
the baryon decay constant from QCD sum rules
and consider different modelizations for the baryon
distribution amplitudes, including the non-relativistic,
QCD sum rules inspired and asymptotic ones.
A lattice calculation of the baryon decay constant in
the nucleon case is consistent, within the inherent
systematic uncertainties of the models,
with that of QCD sum rules \cite{mart89}.
In this context, the non--relativistic $DA$
seems to systematically underestimate
absolute quantities ({\it i.e.}, not
obtained as ratios of amplitudes) like the
decay widths for charmonium exclusive decays, or the hadron
form factors, by two or even three orders of
magnitude, when compared to available experimental results
(see, {\it e.g.}, ref.~\cite{cern89}).
It is then quite reasonable to expect that the use
of more refined distribution amplitudes could lead to
significant modifications in the predicted values
for $a_{_B}$.

There is also a more specific reason to believe that
the dependence of $a_{_B}$ on the $DA$'s
may be not negligible: the expression of $a_{_B}$
is given as a function of squared helicity amplitudes with
different values of the constituent helicities.
Then, the use of distribution amplitudes which,
like those derived from QCD sum rules,
seem to indicate an unusual sharing of the hadron
momentum among its constituents, can substantially modify
different helicity amplitudes.

Based on these motivations, in the rest of this paper
we shall present a new derivation of the parameter
$a_{_B}$, generalizing the results of
ref.~\cite{cari87} to the case of a generic hadron distribution
amplitude. The only restriction on the $DA$ is that
it must satisfy general symmetry properties.
This allows us to compare the results obtained using
different $DA$'s and possibly to get useful informations on them,
independently of the way the baryon decay constant has
been fixed.

We wish also to recall that an alternative approach
for the calculation of $a_{_B}$ has been proposed
by Parisi and Kada.
In ref. \cite{pari92} these authors evaluated
the strong contribution to $a_{_B}$ for
the octet baryons, in the framework of
a quark-(scalar)diquark model for the baryon structure.
Diquarks are another possible way of taking into account
higher order corrections (in particular correlations
between two valence quarks in the baryon). They have been
applied, with good success, to several exclusive processes
at intermediate values of $Q^2$ \cite{anse93c}.

It was stressed by Carimalo
and by Glashow {\it et al.} \cite{glas82}
and will also be argued in the following, that
electromagnetic (e.m.) corrections to $a_{_B}$ might
be by no means negligible.

The main problem with the e.m. corrections is that
they involve the e.m. form factors of the octet baryons,
for which calculations including mass corrections
have not yet been done.
Only for the nucleon, in the non relativistic approximation,
there is a theoretical evaluation \cite{sill86}.
In the other cases we are forced
to give for these corrections estimates based on
the available experimental information.

This paper is organized as follows: in Section II we derive
a general expression for the angular distribution of the
baryons produced in $J/\psi \to B\bar B$ decays.
In Section III we concentrate
on the strong contribution to $a_{_B}$, which
controls the $B\bar B$ angular distribution, giving a derivation of
the helicity amplitudes
required for its calculation.
In Section IV we discuss in detail
the results obtained and their dependence from some
subtleties of the models that in general, while not modifying
qualitatively the conclusions, can substantially affect the numerical
results, in particular for absolute quantities like decay widths.
In Section V we give a detailed analysis of electromagnetic
corrections, trying to estimate, when possible, upper bounds on
the consequent overall modification of $a_{_B}$ both from experimental
and theoretical information. Finally, our conclusions and future
perspectives are discussed in Section VI.

\section{Derivation of the baryon angular distribution
         for $\protect\bbox{J/\psi \to B\bar B}$ decays}
\label{absec}

Let us consider a $J/\psi$ particle,
produced in $e^-e^+$ colliders
with unpolarized beams, which
subsequently decays into a baryon--antibaryon pair:

\begin{equation}
e^-e^+ \to J/\psi \to B\bar B \quad .
\label{decay}
\end{equation}

The spin density matrix of the $J/\psi$, in its rest
frame (which is also the c.m frame of colliding beams,
with the electron moving along the positive $z$ direction),
has the following expression (see, {\it e.g.}, ref.~\cite{bour80}):

\begin{equation}
\rho_{_{MM}\prime}(J/\psi) =
\frac{1}{N}  \sum_{\lambda_{e^-},\lambda_{e^+}}
A^e_{_M;\lambda_{e^-},\lambda_{e^+}}
A^{e\,*}_{_M\prime;\lambda_{e^-},\lambda_{e^+}} \quad ,
\label{rhoJ}
\end{equation}

\noindent
where the $A^e$'s are the helicity amplitudes for the process $e^-e^+
\to J/\psi$, $M$ is the $z$ component of the $J/\psi$ total
angular momentum
in its rest frame and $\lambda_{e^-}$, $\lambda_{e^+}$
are the helicities of the electron and the positron,
respectively; $N$ is a normalization factor, such that
${\rm Tr}[\ \rho\ ] = 1$.

It is not difficult to show that:

\begin{eqnarray}
\rho_{_{MM}\prime} & = & 0 \;\;\;\;\; \mbox{\rm if}\;\; M \neq M'
\label{rhomm} \\
\rho_{11} & = & \rho_{-1,-1} =\frac{1}{2} \left(1+2\frac{m_e^2}{M_\psi^2}
\right)^{-1/2} \cong \frac{1}{2}
\label{rho11} \\
\rho_{00} & = & 2\frac{m_e^2}{M_\psi^2}\left(1+2\frac{m_e^2}{M_\psi^2}\right)
^{-1/2} \cong 0 \quad ,
\label{rho00}
\end{eqnarray}

\noindent where $m_e$ and $M_\psi$ are the masses of the electron and the
$J/\psi$, respectively.
As for the second step in our process, the decay $J/\psi \to B\bar B$,
we have the general relation \cite{jauc55}:

\begin{eqnarray}
 d\Gamma(J/\psi \to B\bar B) & = &
\frac{1}{8(2\pi)^5}\left(1-4\frac{m_{_B}^2}{M_\psi^2}\right)^{1/2} \nonumber \\
& \times & \sum_{M,\lambda_{_B},\lambda_{_{\bar B}}}\,\rho_{_{MM}}
\,|A_{\lambda_{_B}\lambda_{_{\bar B}};M}|^2 d\Omega_{_B} \quad ,
\label{gammabb}
\end{eqnarray}

\noindent where the $A$'s are the helicity amplitudes for the
decay of a $J/\psi$, with third component $M$ of the total angular
momentum $J=1$, into a baryon--antibaryon pair with
helicities $\lambda_{_B}$ and $\lambda_{_{\bar B}}$, respectively;
$m_{_B}$ is the mass of the produced baryons.

Due to the symmetry around the $\hat{z}$ axis, we can
put $\varphi_{_B} = 0$ in our calculations and integrating
over $\varphi_{_B}$ we obtain:

\begin{eqnarray}
\frac{d\Gamma(J/\psi \to B\bar B)}{d(\cos\theta_{_B})} & = &
\frac{1}{8(2\pi)^4}\left(1-4\frac{m_{_B}^2}{M_\psi^2}\right)^{1/2} \nonumber \\
& \times & \sum_{M,\lambda_{_B},\lambda_{_{\bar B}}}\,\rho_{_{MM}}
\,|A_{\lambda_{_B}\lambda_{_{\bar B}};M}|^2 \quad .
\label{dgdcos}
\end{eqnarray}

Apart from overall factors, independent of $\theta_{_B}$,
this is the quantity measured by the MARKII and DM2
collaborations \cite{mark84,dm287}.

We know from first principles \cite{bour80} that the amplitudes
$A_{\lambda_B\lambda_{\bar B};M}$ have
the following general structure

\begin{equation}
A_{\lambda_{_B}\lambda_{_{\bar B}};_M}(\theta_{_B},\varphi_{_B}) =
\tilde A_{\lambda_{_B}\lambda_{_{\bar B}}}\,
d^{\ 1}_{_M,\lambda_{_B}-\lambda_{_{\bar B}}}(\theta_{_B})
\,\exp(iM\varphi_{_B}) \; ,
\label{ampli}
\end{equation}

\noindent where the ``reduced'' amplitude
$\tilde A_{\lambda_{_B}\lambda_{_{\bar B}}}$
is independent of $M$ and the angular
variables and the $d^{\ J}(\theta_{_B})$ are the usual rotation
matrices.
Using the parity properties \cite{bour80} for the
$A_{\lambda_{_B}\lambda_{_{\bar B}};_M}$ (which imply that
$\tilde A_{-+} = \tilde A_{+-}$,
$\tilde A_{--} = \tilde A_{++}$) and for the spin
density matrix $\rho(J/\psi)$ ($\rho_{-1,-1} = \rho_{1,1}$),
Eq.~(\ref{dgdcos}) may be rewritten as follows:

\begin{eqnarray}
\lefteqn{\frac{d\Gamma(J/\psi \to B\bar B)}{d(\cos\theta_{_B})} =
\frac{1}{8(2\pi)^4}\left(1-4\frac{m_{_B}^2}{M_\psi^2}\right)^{1/2} }
\nonumber \\
& & \times \left\{ |\tilde A_{+-}|^2(\rho_{11} + \rho_{00})
+ 2|\tilde A_{++}|^2\rho_{11} \right\} \nonumber \\
& & \times \left\{ 1 + a_{_B}\cos^2\theta_{_B} \right\} \quad ,
\label{dgab}
\end{eqnarray}

\noindent where

\begin{equation}
a_{_B} = \frac{ \left\{ |\tilde A_{+-}|^2 - 2|\tilde A_{++}|^2
\right\}(\rho_{11}-\rho_{00}) }
{ |\tilde A_{+-}|^2(\rho_{11} + \rho_{00})
+ 2|\tilde A_{++}|^2\rho_{11} } \quad .
\label{new}
\end{equation}

If, with good approximation (see Eqs.~(\ref{rho11}),(\ref{rho00})),
we take $\rho_{11} = 1/2$ and $\rho_{00} = 0$, we finally get
the simplified expression

\begin{eqnarray}
\lefteqn{\frac{d\Gamma(J/\psi \to B\bar B)}{d(\cos\theta_{_B})} =
\frac{1}{16(2\pi)^4}\left(1-4\frac{m_{_B}^2}{M_\psi^2}\right)^{1/2} }
\nonumber \\
& & \times \left\{ |\tilde A_{+-}|^2
+ 2|\tilde A_{++}|^2\right\}
\ \left\{ 1 + a_{_B}\cos^2\theta_{_B} \right\} \quad ,
\label{dgab2}
\end{eqnarray}

\noindent where

\begin{equation}
a_{_B} = \frac{ |\tilde A_{+-}|^2 - 2|\tilde A_{++}|^2 }
{ |\tilde A_{+-}|^2 + 2|\tilde A_{++}|^2 } \quad .
\label{ab2}
\end{equation}

We stress again that the results of Eqs.~(\ref{dgab2}),(\ref{ab2})
only require the assumptions that the
one--virtual photon interaction dominates the $J/\psi$
production process and that $m_e^2/M_\psi^2 \cong 0$.

Two further remarks are appropriate: {\it i}) As it is clear
from Eq.~(\ref{ampli}) and from the properties of the
$d^{\ J}$,

\begin{equation}
\tilde A_{+-} = A_{+-;1}(\theta_{_B}=
\varphi_{_B}=0) \quad ,
\label{atfroma1}
\end{equation}

\noindent and

\begin{equation}
\tilde A_{++} = A_{++;0}(\theta_{_B}=
\varphi_{_B}=0) \quad .
\label{atfroma2}
\end{equation}

Then it is sufficient to calculate
the amplitudes $A_{+-;1}$, $A_{++;0}$ in the particular, convenient
kinematic configuration $\theta_{_B} = \varphi_{_B} = 0$,
in order to know, with the help of Eq.~(\ref{ampli}) and of the parity
symmetry properties, all the
amplitudes. {\it ii}) As it will be explicitly shown in the next
Section, if $m_{_B}=0$ then $\tilde A_{++} = 0$ also; in this case
we recover the old PQCD result of Brodsky and Lepage \cite{brod81},
that is $a_{_B} = 1$.

{}From Eq.~(\ref{gammabb}) we easily get the expression of the
total decay width for the process:

\begin{eqnarray}
\Gamma(J/\psi \to B\bar B) & = &
\frac{1}{6(2\pi)^4}\left(1-4\frac{m_{_B}^2}{M_\psi^2}\right)^{1/2} \nonumber \\
& \times & \left\{ |\tilde A_{+-}|^2 + |\tilde A_{++}|^2 \right\} \quad .
\label{gtot}
\end{eqnarray}

\section{The strong contribution to the parameter
\protect$\bbox{\lowercase{a}_{_B}}$: evaluation}
\label{asbsec}

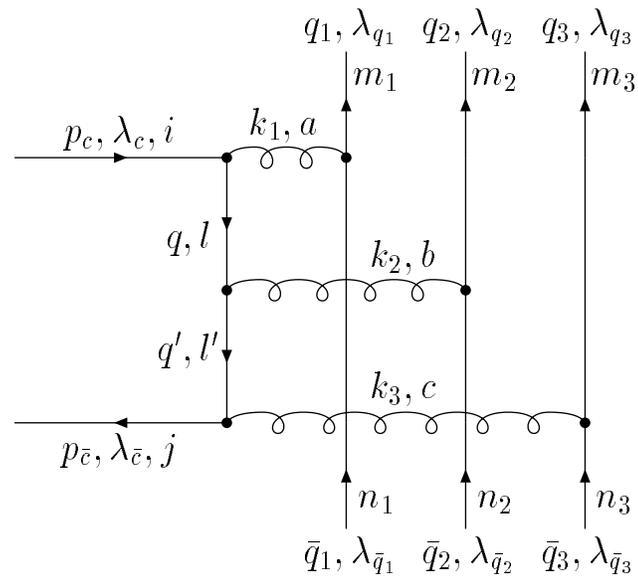
\begin{figure}
\caption{ The Feynman diagram which, to lowest order
in $\alpha_s$, describes the elementary process $Q\bar Q \to
q_1q_2q_3\bar q_1\bar q_2\bar q_3$, for a quarkonium state with
charge conjugation $C=-1$.
In the $Q\bar Q$ center-of-mass
frame, $c^\mu = (E, \bbox{k}/2)$ and
$\bar c^\mu = (E,-\bbox{k}/2)$, where
$\bbox{k}$ is the relative momentum between
the $c$ and $\bar c$ quarks;
$q_{i}=x_{i}p_{B}$ and $\bar q_i=y_ip_{\bar B}$ ($i=1,2,3$),
with $p_B^\mu=(E,\bbox{p}_B)$,
$p_{\bar B}^\mu=(E,-\bbox{p}_B)$ and
$\bbox{p}_B=
(p\sin\theta_B\cos\varphi_B,
p\sin\theta_B\sin\varphi_B, p\cos\theta_B)$.
$a,b,c$, $i,j,l,l',m_{1,2,3}$, $n_{1,2,3}$ are color indices;
the $\lambda$'s label helicities.  }
\label{qcd}
\end{figure}

In this Section and in the next one we shall neglect the
electromagnetic corrections to $a_{_B}$, which will be considered
in Section \ref{aebsec}, and concentrate on the strong
contribution, that from now on will be called $a^s_{_B}$.
As we briefly sketched in Section I, the calculation of the
helicity amplitudes for the physical process $J/\psi \to B\bar B$,
$A^s_{\lambda_{_B}\lambda_{_{\bar B}};M}$
(the suffix $s$ is a reminder that we
are considering only  the strong interaction in what follows),
consists of several steps. First of all we need to calculate
the amplitude for the hard interaction among the elementary (valence)
constituents of the involved hadrons.
To lowest order in the strong coupling constant,
the only (topologically distinct) Feynman graph
is shown in Fig.~\ref{qcd} (where the
notation is also defined). All other possible
graphs of the same order can be obtained from this one by
a permutation of the final fermionic lines. We do not take
into account explicitly all these graphs because
their contribution
is accounted for by opportunely choosing the final hadron
wave functions.
Without giving inessential details of the intermediate steps
of the calculation, we present directly the expression of
this amplitude in the particular kinematic configuration
$\theta_B = \varphi_B = 0$
(the relative momentum between the $c$
and $\bar c$ quarks, $\bbox{k}$, has also been set equal to zero.
This procedure is proper when considering
$L=0$ bound states; for $L \neq 0$ the limit $\bbox{k} \to 0$
should be taken in a subsequent step, {\it i.e.}
after the integration over the angular part of the
charmonium wave function).

\widetext

\begin{eqnarray}
\lefteqn{ T^s_{\lambda_{q_1}\lambda_{q_2}\lambda_{q_3},
\lambda_{\bar q_1}\lambda_{\bar q_2}\lambda_{\bar q_3};
\lambda_c\lambda_{\bar c}}
(\bbox{k}=0;\theta_B = \varphi_B = 0) =
- 16 c_{_F} g_s^6 \frac{1}{M_\psi^5}
\frac{1}{\prod_{i=1}^3 \left[ x_iy_i+(x_i-y_i)^2\epsilon^2_{_B}
\right]} }
\nonumber \\
& & \times \frac{1}{2x_1y_1-x_1-y_1+2(x_1-y_1)^2\epsilon^2_{_B}}
\frac{1}{2x_3y_3-x_3-y_3+2(x_3-y_3)^2\epsilon^2_{_B}} \nonumber \\
& & \times \left\{\ \left[ x_1y_3+x_3y_1+2(x_1-y_1)
(x_3-y_3)\epsilon^2_{_B}
\right] \left[
\delta_{\lambda_{q_1},-\lambda_{\bar q_1}}
\delta_{\lambda_{q_2},-\lambda_{\bar q_2}}
\delta_{\lambda_{q_3},-\lambda_{\bar q_3}}
\delta_{\lambda_{q_1},-\lambda_{q_2}}
\delta_{\lambda_{q_2},-\lambda_{q_3}}
\delta_{\lambda_{q_3},\lambda_{c}}
\delta_{\lambda_{c},-\lambda_{\bar c}} \right.\right. \nonumber \\
& & +\ \epsilon_{_B}
\left(\delta_{\lambda_{q_1},\lambda_{\bar q_1}}
\delta_{\lambda_{q_2},-\lambda_{\bar q_2}}
\delta_{\lambda_{q_3},-\lambda_{\bar q_3}}
\delta_{\lambda_{q_2},\lambda_{c}}
\delta_{\lambda_{q_3},-\lambda_{c}}
\delta_{\lambda_{c},\lambda_{\bar c}}
+ \delta_{\lambda_{q_1},-\lambda_{\bar q_1}}
\delta_{\lambda_{q_2},-\lambda_{\bar q_2}}
\delta_{\lambda_{q_3},\lambda_{\bar q_3}}
\delta_{\lambda_{q_1},\lambda_{c}}
\delta_{\lambda_{q_2},-\lambda_{c}}
\delta_{\lambda_{c},\lambda_{\bar c}} \right)\nonumber \\
& & + \left. \epsilon^2_{_B}
\delta_{\lambda_{q_1},\lambda_{\bar q_1}}
\delta_{\lambda_{q_2},-\lambda_{\bar q_2}}
\delta_{\lambda_{q_3},\lambda_{\bar q_3}}
\delta_{\lambda_{q_2},\lambda_{c}}
\delta_{\lambda_{c},-\lambda_{\bar c}} \right]
- \left[ x_1x_3+y_1y_3-2(x_1-y_1)
(x_3-y_3)\epsilon^2_{_B} \right] \nonumber \\
& & \times \left[ \epsilon_{_B}
\delta_{\lambda_{q_1},-\lambda_{\bar q_1}}
\delta_{\lambda_{q_2},\lambda_{\bar q_2}}
\delta_{\lambda_{q_3},-\lambda_{\bar q_3}}
\delta_{\lambda_{q_1},\lambda_{c}}
\delta_{\lambda_{q_3},-\lambda_{c}}
\delta_{\lambda_{c},\lambda_{\bar c}}
+ \epsilon^2_{_B} \left(
\delta_{\lambda_{q_1},\lambda_{\bar q_1}}
\delta_{\lambda_{q_2},\lambda_{\bar q_2}}
\delta_{\lambda_{q_3},-\lambda_{\bar q_3}}
\delta_{\lambda_{q_3},\lambda_{c}}
\delta_{\lambda_{c},-\lambda_{\bar c}}
\right.\right.\nonumber \\
& & + \left.\left.\left.
\delta_{\lambda_{q_1},-\lambda_{\bar q_1}}
\delta_{\lambda_{q_2},\lambda_{\bar q_2}}
\delta_{\lambda_{q_3},\lambda_{\bar q_3}}
\delta_{\lambda_{q_1},\lambda_{c}}
\delta_{\lambda_{c},-\lambda_{\bar c}} \right)
+ \epsilon^3_{_B}
\delta_{\lambda_{q_1},\lambda_{\bar q_1}}
\delta_{\lambda_{q_2},\lambda_{\bar q_2}}
\delta_{\lambda_{q_3},\lambda_{\bar q_3}}
\delta_{\lambda_{c},\lambda_{\bar c}}
\right] \right\}
\label{T}
\end{eqnarray}

\noindent where $c_{_F}$ is the color factor which, once the convolution
with the final hadron wave functions is made, takes the value
$c_{_F} = 5/(18\sqrt{3})$ and, as usual, $g_s = \sqrt{4\pi\alpha_s}$;
$x_i(y_i)$ represents the (light-cone) fraction of the baryon (antibaryon)
four--momentum carried by the $i$--th quark (antiquark);
$\epsilon_{_B} = m_{_B}/M_\psi$.
It is clear from this equation the role played by mass corrections
in allowing spin--flips along the final fermionic lines.
We see that higher $\epsilon_{_B}$ powers correspond to terms where
more spin--flips are present. On the contrary,
if we neglect quark masses Eq.~(\ref{T}) reduces, apart
from an inessential constant factor due to a different notation,
to the result of Brodsky and Lepage \cite{brod81}:

\begin{eqnarray}
\lefteqn{ T^{(0)s}_{\lambda_{q_1}\lambda_{q_2}\lambda_{q_3},
\lambda_{\bar q_1}\lambda_{\bar q_2}\lambda_{\bar q_3};
\lambda_c\lambda_{\bar c}} =
- 16 c_{_F} g_s^6 \frac{1}{M_\psi^5}
\frac{1}{\prod_{i=1}^3 \ x_iy_i} \frac{1}{2x_1y_1-x_1-y_1}
\frac{1}{2x_3y_3-x_3-y_3} } \nonumber \\
& & \times (x_1y_3+x_3y_1)
\delta_{\lambda_{q_1},-\lambda_{\bar q_1}}
\delta_{\lambda_{q_2},-\lambda_{\bar q_2}}
\delta_{\lambda_{q_3},-\lambda_{\bar q_3}}
\delta_{\lambda_{q_1},-\lambda_{q_2}}
\delta_{\lambda_{q_2},-\lambda_{q_3}}
\delta_{\lambda_{q_3},\lambda_{c}}
\delta_{\lambda_{c},-\lambda_{\bar c}} \quad .
\label{T0}
\end{eqnarray}

\narrowtext

We would like to stress that, in principle, the $g_s^2$ factors
coming from different virtual gluons should be evaluated at the
corresponding values of transferred $Q^2$ \cite{brod89}. For the moment
we do not discuss explicitly this problem, which will
be analyzed in detail in the next Section, where numerical results
are presented.

Next we convolute
the elementary helicity amplitude $T^s_{\{\lambda\}}$
(for brevity we indicate by $\{\lambda\}$ the collection of
all the helicities from which an amplitude depends)
with the final hadron wave functions:

\begin{equation}
M^s_{\lambda_{_B}\lambda_{_{\bar B}};\lambda_c\lambda_{\bar c}} =
\int\ [d\tilde x][d\tilde y]
\psi_{_B,\lambda_{_B}}
(\tilde x)T^s_{\{\lambda\}}(\tilde x,\tilde y)
\psi_{_{\bar B},\lambda_{_{\bar B}}}(\tilde y) \; ,
\label{M}
\end{equation}

\noindent where $\int [d\tilde z]$ stays for $\int_0^1 dz_1dz_2dz_3
\delta(1-z_1-z_2-z_3)$ and we used $\tilde z$
as a shorthand for $(z_1,z_2,z_3)$. The amplitude
$M^s_{\{\lambda\}}$  refers to the decay of a free $c$, $\bar c$
quark pair into the final baryon--antibaryon pair.

Once the amplitudes $M^s$ have been evaluated (we will do
this below), the final step consists in integrating these
amplitudes over the proper $c\bar c$--bound state wave function,
taken as usual in the non--relativistic approximation, such that
the physical amplitude $A^s_{\lambda_{_B},\lambda_{_{\bar B}};M}$,
to which we are interested in, is given by the following general expression:

\begin{eqnarray}
\lefteqn{ A^s_{\lambda_{_B},\lambda_{_{\bar B}};M} =
\sum_{\lambda_c\lambda_{\bar c}} \left(
{2L+1 \over 4\pi} \right)^{1/2}
C^{{1\over2}\;\;{1\over2}\;\;S}_{\lambda_c,\,-\lambda_{\bar c}
\:\lambda}\,C^{L\, S\, J}_{0\:\lambda\:\lambda} } \nonumber \\
& & \times \int d^3k \,M^s_{\lambda_{_B},\lambda_{_{\bar B}};
\lambda_c\lambda_{\bar c}}(\bbox{k})
D^{J\,*}_{M\lambda}(\beta,\alpha,0)
\,\psi_C(k) \quad ,
\label{ageneral}
\end{eqnarray}

\noindent where the $C$'s are the
Clebsh--Gordan coefficients,
$\lambda = \lambda_c - \lambda_{\bar c}$,
$\bbox{k} = (k,\alpha,\beta)$ is
the relative momentum between the $c$ and $\bar c $ quarks
and finally $\psi_c(k)$ is the (momentum--space) charmonium
wave function.
In particular, for the $J/\psi$ $L=0$, so we can take
from the beginning (in the full non--relativistic approximation)
$\bbox{k} \to 0$
without loss of generality (this property
has been used in the derivation of Eq.~(\ref{T})).
Then we can see that:

\begin{equation}
A^s_{+-;1} = \sqrt{2}\ \pi|R_s(0)|M^s_{+-;+-} \quad ,
\label{A1}
\end{equation}

\begin{equation}
A^s_{++;0} = 2\pi|R_s(0)|M^s_{++;++} \quad ,
\label{A2}
\end{equation}

\noindent where $R_s(0)$ is the value of the $L = 0$ charmonium wave function
at the origin.
By comparing the theoretical prediction for the decay width
$\Gamma(J/\psi \to e^-e^+)$, $\Gamma_{ee} \cong (16/9)\alpha^2
|R_s(0)|^2/M_{\psi}^2$, with the available experimental
data \cite{pdg92} we can estimate $|R_s(0)| \simeq 0.737$ GeV$^{3/2}$.
Here we have considered only the two amplitudes that, as it
was discussed in the previous Section, are sufficient to
recover the expressions of all the others, when use is made
of Eq.~(\ref{ampli}). In Eq.~(\ref{A2}) the relation
$M^s_{++;--} = M^s_{++;++}$ has also been used (the validity of
this relation can be proved from Eqs.~(\ref{T}),(\ref{M})).

As it is clear from Eq.~(\ref{M}), while the $T^s$ amplitude is
the same for all the baryon pairs considered, the $M^s$ amplitudes
are different for the different baryons, essentially
because in general the spin--flavor component of their
wave function changes. So, we cannot give a general, explicit
expression for the amplitude $M^s$, but we must consider separately
all the different cases. The total hadron wave function consists
of a color part (which is the same for all the baryons
and has been englobed in the definition of the color factor
of the amplitude $T^s$), a spin--flavor component and a dynamical
part which describes, in momentum space, how the baryon
four--momentum is shared among its valence constituents: the
distribution amplitude. In general the
$DA$, which as we said previously is a non--perturbative quantity,
allows for non $SU_f(3)$-symmetric configurations, as it has been shown
by several studies performed with the help of QCD sum--rules
techniques \cite{cern89,cern84,cer84b,king87,gari87,stef92}.
However, in the particular case of an
$SU_f(3)$-symmetric DA, as we shall
see, we obtain the same result for all the baryons and the
only differences are the (experimental) values of the
baryon masses \cite{cari87}.

In order to explicitly calculate the $M$ amplitudes
and their dependence from the $DA$'s
we need to take from the literature the available proposed
models.
QCD sum--rules results for octet baryons $DA$'s exist at present only
for the nucleon
\cite{cern89,cern84,cer84b,king87,gari87,stef92}
and for the $\Sigma^+$, $\Xi^-$ and $\Lambda$ \cite{cern89}.
Then, below we give explicitly the expression for the distribution
amplitudes and for the required $M^s$ amplitudes in these
cases.

The most general wave functions for the baryons considered here
are the following:

\noindent for $p$, $n$, $\Sigma^+$, $\Xi^-$ baryons:

\widetext

\begin{eqnarray}
\psi_{B, \lambda_{B}}(\tilde{x}) & = & 2\lambda_{B}
\frac{F_{B}}{4\sqrt{6}}\{\varphi_{B}(123)f_{1,\lambda_{B}}(1)
f_{2,-\lambda_{B}}(2)f_{3,\lambda_{B}}(3)
+ \varphi_{B}(213)f_{1,-\lambda_{B}}(1)f_{2,\lambda_{B}}(2)
f_{3,\lambda_{B}}(3) \nonumber \\
& - &  2R_B T_{B}(123)f_{1,\lambda_{B}}(1)f_{2,\lambda_{B}}(2)
f_{3,-\lambda_{B}}(3)
+ (1 \longleftrightarrow 3) + (2 \longleftrightarrow 3)\ \} \; ;
\label{psib}
\end{eqnarray}

\noindent for the $\Lambda$ baryon:

\begin{eqnarray}
\psi_{\Lambda, \lambda_{\Lambda}}(\tilde{x}) & = & 2\lambda_{\Lambda}
\frac{F_{\Lambda}}{4\sqrt{6}}\{\varphi_{\Lambda}(123)
u_{\lambda_{\Lambda}}(1)
d_{-\lambda_{\Lambda}}(2)s_{\lambda_{\Lambda}}(3)
- \varphi_{\Lambda}(213)u_{-\lambda_{\Lambda}}(1)
d_{\lambda_{\Lambda}}(2)
s_{\lambda_{\Lambda}}(3) \nonumber \\
& - & 2R_{\Lambda}T_{\Lambda}(123)u_{\lambda_{\Lambda}}(1)
d_{\lambda_{\Lambda}}(2)
s_{-\lambda_{\Lambda}}(3)
+ \;\; \mbox{\rm all the permutations of} \; (1, \; 2, \; 3)\  \} \; .
\label{psilambda}
\end{eqnarray}

\noindent We have introduced the notation

\begin{equation}
R_B = \frac{F^T_B}{F_B} \quad ,
\label{Rt}
\end{equation}

\noindent where $F_B$ and $F_B^T$ are constants related to the value of
the baryon wave function at the origin.
In Eq.~(\ref{psib}) $f_{1,2,3}$ are the flavors
appropriate to the particular baryon considered
($f_{1,2,3}=uud$ for the proton, $udd$ for the neutron,
$uus$ for the $\Sigma^+$ and $ssd$ for the $\Xi^-$).
Isospin symmetry properties impose several relations
between $\varphi_{_B}(\tilde x)$ and $T_{_B}(\tilde x)$;
in the case of the nucleon the relations $R_N = 1$,
$2T_N(1,2,3) = \varphi_{N}(1,3,2)+\varphi_{N}(2,3,1)$
also hold (see ref. \cite{cern89} for further details).
By insertion of these expressions and Eq.~(\ref{T})
in Eq.~(\ref{M}), we can, after some algebra,
derive the following $M^s$ amplitudes.

\noindent For $N$, $\Sigma^+$, $\Xi^-$ baryons:
\begin{eqnarray}
M^s_{++;++} & = &
\frac{F_{B}^2}{96}\epsilon_{_B}\int[d\tilde{x}][d\tilde{y}]C_{B}
\biggl\{2D_{B}\Bigl[\varphi_{B}(123)\varphi_{B}(213)
-2R_{B}\mbox{\large(}\varphi_{B}(321)T_{B}(321) \nonumber \\
& & +\ \varphi_{B}(312)T_{B}(132)\mbox{\large)}\Bigr]
+ E_{B}\Bigl[\varphi_{B}(132)\varphi_{B}(312) -
4R_{B}T_{B}(123)\varphi_{B}(213)\Bigr] \nonumber \\
& & +\ 2E_{B}\epsilon^2_{_B}\Bigl[\varphi_{B}^{2}(123)
+ \varphi_{B}^{2}(213) +
\varphi_{B}^{2}(312)
+ 2R_{B}^{2}\mbox{\large(}2T_{B}^{2}(123) + T_{B}^{2}(132)\mbox{\large)}
\Bigr]\biggr\} \; ,
\label{Mbpppp}
\end{eqnarray}

\begin{eqnarray}
M^s_{+-;+-} & = &
-\frac{F_{B}^2}{48}\int[d\tilde{x}][d\tilde{y}]C_{B}
\biggl\{D_{B}\Bigl[\varphi_{B}^{2}(123) +
2R_{B}^{2}T_{B}^{2}(132)\Bigr] \nonumber \\
& & +\ D_{B}\epsilon^2_{_B}\Bigl[\varphi_{B}(132)
\varphi_{B}(312) -
4R_{B}T_{B}(123)\varphi_{B}(213)\Bigr] + 2E_{B}\epsilon^2_{_B}\Bigl[
\varphi_{B}(123)\varphi_{B}(213) \nonumber \\
& & -\ R_{B}\mbox{\large(}2\varphi_{B}(123)T_{B}(123)
+ \varphi_{B}(312)T_{B}(132)+\varphi_{B}(132)
T_{B}(132)\mbox{\large)}\Bigr]\biggr\} \; .
\label{Mbpmpm}
\end{eqnarray}

\noindent For the $\Lambda$ baryon:

\begin{eqnarray}
M^s_{++;++} & = &
\frac{F_{\Lambda}^2}{48}\epsilon_{_\Lambda}
\int[d\tilde{x}][d\tilde{y}]C_{\Lambda}
\biggl\{-2D_{\Lambda}\Bigl[\varphi_{\Lambda}(123)
\varphi_{\Lambda}(213)
+ 2R_{\Lambda}\mbox{\large(}\varphi_{\Lambda}(321)T_{\Lambda}(321) \nonumber \\
& & +\ \varphi_{\Lambda}(312)T_{\Lambda}(312)\mbox{\large)}\Bigr]
- E_{\Lambda}\Bigl[\varphi_{\Lambda}(132)
\varphi_{\Lambda}(312) +
4R_{\Lambda}\varphi_{\Lambda}(213)T_{\Lambda}(213)\Bigr] \nonumber \\
& & +\ 2E_{\Lambda}\epsilon^2_{_\Lambda}\Bigl[\varphi_{\Lambda}^{2}(123)
+ \varphi_{\Lambda}^{2}(213) +
\varphi_{\Lambda}^{2}(312)
+ 2R_{\Lambda}^{2}\mbox{\large(}2T_{\Lambda}^{2}(123)
+ T_{\Lambda}^{2}(312)\mbox{\large)}\Bigr]\biggr\} \; ,
\label{Mlampppp}
\end{eqnarray}

\begin{eqnarray}
M^
s_{+-;+-} & = &
-\frac{F_{\Lambda}^2}{24}\int[d\tilde{x}][d\tilde{y}]C_{\Lambda}
\biggl\{D_{\Lambda}\Bigl[\varphi_{\Lambda}^{2}(123) +
2R_{\Lambda}^{2}T_{\Lambda}^{2}(132)\Bigl] \nonumber \\
& & -\ D_{\Lambda}\epsilon^2_{_\Lambda}\Bigl[\varphi_{\Lambda}(312)
\varphi_{\Lambda}(132) -
4R_{\Lambda}\varphi_{\Lambda}(213)T_{\Lambda}(123)\Bigr] \nonumber \\
& & -\ 2E_{\Lambda}\epsilon^2_{_\Lambda}\Bigl[\varphi_{\Lambda}(123)
\varphi_{\Lambda}(213) +
2R_{\Lambda}\mbox{\large(}\varphi_{\Lambda}(321)T_{\Lambda}(321)
+ \varphi_{\Lambda}(312)T_{\Lambda}(312)\mbox{\large)}\Bigr]\biggr\} \; .
\label{Mlampmpm}
\end{eqnarray}

\narrowtext

In Eqs.~(\ref{Mbpppp})--(\ref{Mlampmpm}) the following concise
notation has been used: in each $\varphi(i,j,k)\varphi(l,m,n)$,
$\varphi(i,j,k)T(l,m,n)$ and
$\varphi^2(i,j,k)$=$\varphi(i,j,k)\varphi(i,j,k)$
product the first term is a function
of $\tilde x$, the second of $\tilde y$.
Furthermore we have defined:

\begin{equation}
D_{B} = x_{1}y_{3} + x_{3}y_{1} + 2(x_{1} - y_{1})
(x_{3} - y_{3})\epsilon^2_{_B}
\label{Db}
\end{equation}

\begin{equation}
E_{B} =  - (x_{1}x_{3} + y_{1}y_{3})
         + 2(x_{1} - y_{1})(x_{3} - y_{3})\epsilon^2_{_B}
\label{eb}
\end{equation}

\begin{eqnarray}
 C_{B} & = & -\frac{2560\pi^{3}\alpha_{s}^{3}}{9\sqrt{3}M_\psi^{5}}
\frac{1}{\prod_{i=1}^{3}[x_{i}y_{i} + (x_{i} - y_{i})^{2}
\epsilon^2_{_B}]}
\nonumber \\
& \times & \frac{1}{[2x_{1}y_{1} - x_{1} - y_{1}
+ 2(x_{1} - y_{1})^{2}\epsilon^2_{_B}]} \nonumber \\
& \times & \frac{1}{[2x_{3}y_{3} - x_{3} - y_{3}
+ 2(x_{3} - y_{3})^{2}\epsilon^2_{_B}]}
\label{Cb}
\end{eqnarray}

Let us finally stress that, as it was anticipated, all the
$M^s_{++;++}$ amplitudes vanish if we take $m_{_B} \to 0$
(this means that $a^s_{_B}=1$),
as it must be, because in this case there is not in the model
any mechanism which allows for spin flips in the
quark--gluon vertices, thus forcing the final baryons to have
opposite helicities.

\section{The strong contribution to the parameter
\protect$\bbox{\lowercase{a}_{_B}}$: numerical results}
\label{asbris}

In the expressions presented in the previous Section for the
$M^s$ amplitudes, from which $a^s_{_B}$ can be evaluated
(see Eqs.~(\ref{ab2}),(\ref{T}),(\ref{M}),(\ref{A1}),(\ref{A2}))
we are left with the
unknown form of the distribution amplitudes, which
contain all the hadronization dynamics of the
partons of the dominant, valence light--cone Fock state.
As we said previously, these $DA$'s are highly non--perturbative
in nature, so perturbative QCD cannot say about them much
more than their general, formal solution (expressed
as an infinite sum of Appell polynomials with
unknown, non--perturbative, coefficients) and their
evolution in $Q^2$ \cite{brod89}. This is by no means a trivial
result, but it is not enough to give us reliable expressions
for the $DA$ at values of $Q^2$ presently accessible.
For $Q^2 \to \infty$, only the first term in the formal
expansion of the wave function survives, so we obtain a simple
expression, the so called asymptotic distribution
amplitude. The asymptotic $DA$ again contains an unknown coefficient,
which however may be extracted from the best experimental data
at hand.
The present status of non--perturbative methods
(namely QCD sum rules and Lattice calculations, in our specific case)
allows us to calculate only  a few low moments of the $DA$'s.
{}From these, a model expression for the $DA$ is proposed
by opportunely truncating the infinite, formal expansion
of perturbative QCD and finding approximate values of the
unknown expansion coefficients by fitting the calculated
moments. There are, of course, several difficulties in this
procedure. These are mainly reflected in an apparently strong
dependence of the model $DA$ from the number of available moments;
in sizeable differences among the
$DA$'s proposed by different groups; in discrepancies between the
QCD sum rules and the Lattice results.
In our calculation, we will take a phenomenological approach,
considering all the $DA$ expressions available and analyzing
how the results for $a^s_{_B}$ depend on each of them.

Let us first consider the case of the nucleon. There are
several versions of QCD sum rules model $DA$'s in the
literature. For completeness we give these below, together
with the non--relativistic and the asymptotic $DA$'s.

\begin{equation}
\varphi^{NR}(\tilde{x}) = \prod_{i=1}^3
\delta\left(x_{i}-\frac{1}{3}\right) \; ,
\label{phinr}
\end{equation}

\begin{equation}
\varphi^{AS}(\tilde{x}) = 120x_{1}x_{2}x_{3} \; .
\label{phias}
\end{equation}

\noindent The $DA$ proposed by Chernyak and Zithnitsky \cite{cern84}

\begin{eqnarray}
\varphi_{N}^{CZ}(\tilde{x}) & = & \varphi^{AS}(\tilde{x})
[18.06x_{1}^{2}+ 4.62x_{2}^{2} \nonumber \\
& + & 8.82x_{3}^{2} - 1.68x_{3} - 2.94] \; .
\label{phicz}
\end{eqnarray}

\noindent The $DA$ of Chernyak, Ogloblin and Zhitnitsky \cite{cern89}

\begin{eqnarray}
\varphi_{N}^{COZ}(\tilde{x}) & = & \varphi^{AS}(\tilde{x})
[23.814x_{1}^{2} + 12.978x_{2}^{2} \nonumber \\
& + & 6.174x_{3}^{2} + 5.88x_{3} - 7.098] \; .
\label{phicoz}
\end{eqnarray}

\noindent The $DA$ of King and Sachrajda \cite{king87}

\begin{eqnarray}
\varphi_{N}^{KS}(\tilde{x}) & = & \varphi^{AS}(\tilde{x})
[20.16x_{1}^{2} + 15.12x_{2}^{2} + 22.68x_{3}^{2} \nonumber \\
& - & 6.72x_{3} + 1.68(x_{1} - x_{2})- 5.04] \; .
\label{phiks}
\end{eqnarray}

\noindent That of Gari and Stefanis \cite{gari87}

\begin{eqnarray}
\varphi_{N}^{GS}(\tilde{x}) & = & \varphi^{AS}(\tilde{x})
[-1.027x_{1}^{2} + 12.307x_{3}^{2} + 25.88x_{2} \nonumber \\
& + & 111.32x_{1}x_{3} + 9.105(x_{1} - x_{3})- 19.84] \; .
\label{phigs}
\end{eqnarray}

The improved (Heterotic) version of the Gari-Stefanis $DA$,
proposed by Stefanis and Bergmann \cite{stef92}

\begin{eqnarray}
\varphi_{N}^{HET}(\tilde{x}) & = & \varphi^{AS}(\tilde{x})
[- 19.773 + 32.756(x_{1} - x_{3})  \nonumber \\
& + & 26.569x_{2}+16.625x_{1}x_{3} \nonumber \\
& - & 2.916x_{1}^{2} + 75.25x_{3}^{2}] \; .
\label{phihet}
\end{eqnarray}

In all these cases the nucleon decay constant $F_N$ has
roughly the same value:

\begin{equation}
|F_N| \cong 5.0 \cdot 10^{-3}\,\mbox{GeV}^2
\label{fn}
\end{equation}

Furthermore, as we said previously, $R_N = 1$ and
$2\,T_N(1,2,3) = \varphi_{N}(1,3,2)+\varphi_{N}(2,3,1)$
(see Eq.~(\ref{psib})).

The only parameter that must be fixed is the value of
the strong coupling constant $\alpha_s$. For the moment we will limit
ourselves to take a fixed value of $\alpha_s$ for all
three virtual gluons, using the perturbative
expression for the running coupling constant:

\begin{equation}
\alpha_s(Q^2) = \frac{12\pi}{(11n_c-2n_f)\log\left( Q^2/\Lambda^2 \right)}
\quad ,
\label{run}
\end{equation}

\noindent where $n_c=3$ is the number of colors and $n_f$ is
the number of active flavors (in this context, $n_f=4$).
{}From this point of view $Q^2$ must be set to an overall
effective value, pertinent to the process considered, and
we take the square of the $J/\psi$ mass as this effective scale; we also
use $\Lambda \cong 0.2 $ GeV.

This is by no means the only possible way to treat
$\alpha_s$, as we shall see in the following. Of course, these
ambiguities reflect some uncompletness of the theoretical
models.

In order to evaluate the helicity amplitudes
$A_{\lambda_B\lambda_{\bar B};M}$ we finally need to compute
quadruple integrals, which is done numerically.
We have tested both our analytical results (the expressions
of the amplitudes shown in the previous Section) and our numerical
calculations. For example, it is not difficult to see
that if we consider a non--relativistic $DA$ we recover, through
the different steps of our calculation, the analytical
form of $a^s_{_B}$ proposed by Carimalo \cite{cari87}.
As for the numerical integrations, we have explicitly checked that
if one of the usual representations of the Dirac function
is inserted in the integrals, and a limit procedure is made
for the parameter which enter the representation, the
results smoothly tend to the analytical ones in the
non--relativistic case.

In Table~I we give our results for $a^s_{_N}$ and for the
total decay width $\Gamma^s(J/\psi \to N\bar N) = \Gamma^s_{N\bar N}$
for the nucleon. Proton and neutron results differ in this case
only for the small mass difference, so we do not present separate
results (the situation will of course be different when we shall
consider, in the next Section, the electromagnetic corrections).

A comparison of our numerical results for the total decay widths
with those of ref.~\cite{cari87,cern89} needs some care.
Ref.~\cite{cern89} presents results only in the massless
case and using $\alpha_s \simeq 0.3$ and $|R_s(0)| \simeq
0.690$ GeV$^{3/2}$. When opportunely rescaled to these
parameterizations our massless results are in good agreement
with those of ref.~\cite{cern89}.
Note, however, that from an analytical point of view
our expression for $\Gamma(J/\psi \to B\bar B)$ in the
massless case differs from that of ref.~\cite{cern89}
by an overall factor of 80/81.

In the case of the non relativistic $DA$ our analytical
expression of $\Gamma(J/\psi \to B\bar B)$ agrees with
that of ref.~\cite{cari87}.
Let us stress, however, that our
$F_N = 2^{5/2}F_N$(ref.~\cite{cari87});
that is, $F_N$(ref.~\cite{cari87}) $\sim 5\cdot 10^{-3}$
GeV$^2$ corresponds to our $F_N \sim 28\cdot 10^{-3}$ GeV$^2$.

As it can be seen from Table~I, QCD sum rules $DA$'s
give results which are sizably different from those
of the non--relativistic case, even for an observable
like $a^s_{_B}$ which is given as a ratio among squared
helicities amplitudes.
All the results (with the possible exception of the $GS$ $DA$)
are in agreement with the experimental measurements,
but we must not forget that the experimental errors are large
at present, and that electromagnetic contributions can
in principle modify the theoretical values.
We see also that there are non negligible differences
among the various QCD sum rules $DA$'s.
Of course the discrepancies are dramatically
larger in the case of decay widths, so one can ask why study
the (relatively) little  dependence of $a^s_{_B}$ on the $DA$.
The point is that, as we shall see better in the following,
the same reason which makes variations in $a^s_{_B}$ so small
also tend to make $a^s_{_B}$ freer from several ambiguities of
the model, giving, in our opinion, a more sensible
test for the distribution amplitudes.
However, in order to discriminate among different $DA$'s
we need better experimental
results than those presently at disposal.
We hope they will be available in the near future.
Note also that, with $F_N$ given by Eq.~(\ref{fn}),
QCD sum--rules $DA$'s give results for the decay widths
which are in rough agreement with the experiment, unlike the
asymptotic and non--relativistic ones (here, as in the
following, no particular attempt in modifying parameters
in order to better reproduce experimental results has been
made).

Apart from the nucleon, little is known about the form
of the distribution amplitudes of the other octet baryons.
To our knowledge, the only available models are those
proposed by Chernyak {\it et al.} \cite{cern89} for the $\Sigma^+$,
$\Xi^-$ and $\Lambda$.

\noindent From ref. \cite{cern89} we have, for the $\Sigma^+$:

\begin{eqnarray}
\varphi_{\Sigma}^{COZ}(\tilde{x}) & = & 42\varphi^{AS}(\tilde{x})
[0.36x_{1}^{2} +0.24x_{2}^{2} \nonumber \\
& + & 0.14x_{3}^{2} - 0.54x_{1}x_{2} \nonumber \\
& - & 0.16x_{3}(x_{1}+x_{2})+0.05(x_{1}-x_{2})] \quad ,
\label{phisig}
\end{eqnarray}

\begin{eqnarray}
T_{\Sigma}^{COZ}(\tilde{x}) & = & 42\varphi^{AS}(\tilde{x})
[0.32(x_{1}^{2} +x_{2}^{2})+0.16x_{3}^{2} \nonumber \\
& - & 0.47x_{1}x_{2}-0.24x_{3}(x_{1}+x_{2})] \quad .
\label{tsig}
\end{eqnarray}

The decay constants are $|F_\Sigma| \simeq
5.1\cdot 10^{-3}$ GeV$^2$, $|F^T_\Sigma| \simeq
4.9\cdot 10^{-3}$ GeV$^2$.

\noindent For the $\Xi^-$:

\begin{eqnarray}
\varphi_{\Xi}^{COZ}(\tilde{x}) & = & 42\varphi^{AS}(\tilde{x})
[0.38x_{1}^{2}+0.20x_{2}^{2} \nonumber \\
& + & 0.16x_{3}^{2} - 0.26x_{1}x_{2} \nonumber \\
& - & 0.30x_{3}(x_{1}+x_{2})+0.02(x_{1}-x_{2})] \quad ,
\label{phixi}
\end{eqnarray}

\begin{eqnarray}
T_{\Xi}^{COZ}(\tilde{x}) & = & 42\varphi^{AS}(\tilde{x})
[0.28(x_{1}^{2}+x_{2}^{2})+0.18x_{3}^{2} \nonumber \\
& - & 0.16x_{1}x_{2}-0.35x_{3}(x_{1}+x_{2})] \quad ,
\label{txi}
\end{eqnarray}

\noindent with $|F_\Xi| \simeq
5.3\cdot 10^{-3}$ GeV$^2$, $|F^T_\Xi| \simeq
5.4\cdot 10^{-3}$ GeV$^2$.

\noindent Finally, for the $\Lambda$:

\begin{eqnarray}
\varphi_{\Lambda}^{COZ}(\tilde{x}) & = & 42\varphi^{AS}(\tilde{x})
[0.44x_{1}^{2}+0.08x_{2}^{2} \nonumber \\
& + & 0.34x_{3}^{2} - 0.56x_{1}x_{2} \nonumber \\
& - & 0.24x_{3}(x_{1}+x_{2})-0.10(x_{1}-x_{2})] \quad,
\label{phila}
\end{eqnarray}

\begin{equation}
T_{\Lambda}^{COZ}(\tilde{x}) = 42\varphi^{AS}(\tilde{x})[1.2(x_{2}^{2}
-x_{1}^{2})+1.4(x_{1}-x_{2})] \quad ,
\label{tla}
\end{equation}

\noindent and $|F_\Lambda| \simeq
6.3\cdot 10^{-3}$ GeV$^2$, $|F^T_\Lambda| \simeq
6.3\cdot 10^{-4}$ GeV$^2$.

In Table~II, we compare the values of
$a^s_{_B}$ obtained from these QCD sum rules $DA$'s with
those obtained in the non--relativistic case and with
experimental results (note that, in the case of the $\Sigma^+$,
experimental data are available for $\Sigma^0$ only; e.m. corrections can
in principle be different for $\Sigma^+$ and $\Sigma^0$,
so a strict comparison of our results with the experimental
data could be at this stage misleading).
We see from Table~II that for the $\Sigma^+$, $\Xi^-$ and $\Lambda$
differences between non--relativistic and QCD sum--rules
$DA$'s results are also more marked. This is not
unexpected, in that the presence of one or more valence $s$ quarks
breaks more severely the $SU_f(3)$ symmetry implicit in the
$NR$ $DA$.
Again, more precise measurements
of $a_{_B}$ could be very useful in improving our understanding
of (at least) the main features of the $DA$'s.
By the way, we stress that the decay widths for $\Sigma^+$,
$\Xi^-$, $\Lambda$ seem to be poorly reproduced, also
when QCD-sum rules $DA$'s are used. Even if e.m. corrections
are not accounted for, it is unlikely that they can be able
to improve very much the results in this sense.

We want now  to analyze better how our results depend on some
ingredients of the models that, due to their (to some degree)
effective nature, are not unambiguously fixed.

We basically concentrate on the behavior of
the results when different ways of treating the strong coupling
constant are considered. We stress that this is probably
(together with the value of the baryon decay constant $F_B$) the
main source of indeterminacy of the quantitative
results of the model, once the $DA$ has been fixed.

As we anticipated previously, when the calculations of the
results quoted in Table~I and Table~II were discussed,
the definition of
$\alpha_s$ is in some sense ambiguous. We can make
the following (reasonable but not derived from first
principles) choices:

\begin{description}
 \item[{\it i})] Consider an overall, effective
 value of $Q^2$, $Q^2_{e}$, which sets the scale to
which $\alpha_s$ must be evaluated, using Eq.~(\ref{run}),
neglecting that in the model the momenta carried
by the three virtual gluons in the hard scattering
depend on $\tilde x, \tilde y$ and are different
among them \cite{brod89}.
In particular, we choose $Q^2_{e} \sim M_\psi^2$.
Slightly different choices have been used in the literature
(see {\it e.g.} ref.~\cite{cern89,cern84});
this is equivalent to an overall rescaling of
the decay widths, while $a_{_B}$ is unchanged.

\item[{\it ii})] Take a running coupling constant depending
 on the squared momenta carried by the hard gluons,
$\tilde Q_i^2 = Q_i^2(\tilde x, \tilde y)$ (where $i$ runs
over the three gluons) \cite{brod89}.
However, if this choice is made, we face a
serious problem: over the full field of variation of
$\tilde x, \tilde y$, the $\tilde Q^2_i$ approach zero.
There, then, it does not make sense to speak of perturbative calculations,
since $\alpha_s$ increases more and more, invalidating any
perturbative power expansion.
Often people prevents these problems by
introducing an {\it ad hoc} cut--off, $Q^2_0$, such that

\begin{equation}
  \alpha_s(Q^2) = \left\{ \begin{array}{ll}
\displaystyle{ \frac{12\pi}{(11n_c-2n_f)\log\left(Q^2/\Lambda^2\right)} }
& \mbox{if $Q^2 > Q^2_0$} \\
\vbox{\vskip6pt} \\
\alpha_0 = \alpha_s(Q^2_0) & \mbox{if $Q^2 \leq Q^2_0$}
\end{array}
\right.
\label{runcut}
\end{equation}

\noindent where usually one takes $\alpha_0 = 0.3$ or $0.5$.
In some sense, this is a na\"{\i}ve way of
accounting for non--perturbative effects which prevent
$\alpha_s$ to take higher and higher values (like
Eq.~(\ref{run}) should imply), even if probably the exact behavior of
$\alpha_s$ is only poorly mimed.
Even if the $\tilde Q^2_i \to 0$ region is suppressed
by the behavior of the distribution amplitudes,
this procedure is far from being satisfactory.
Furthermore it introduces a dependence from the new parameter $\alpha_0$.

\item[{\it iii})] Take into account in a more
serious way non--perturbative
effects, which may give rise to an
effective mass for the gluons (see, {\it e.g.},
ref.~\cite{corn82}).
This in turn modifies the gluon propagators and the perturbative
expression for $\alpha_s$, Eq.~(\ref{run}).
To be rigorous, there are probably other minor modifications to be
introduced into the fermion propagators
and the quark-gluon vertices as well, if one
consistently applies these methods. However, since we are only trying
to estimate the dependence of our results from different ways of treating
$\alpha_s$, we limit ourselves to consider only the effective
gluon mass effects on it.
In this context, the expression of $\alpha_s$ to be used
is of the type:

\begin{equation}
\alpha_s(Q^2) = \frac{12\pi}
                {(11n_c-2n_f)\log\left[(Q^2+4m_g^2)/\Lambda^2\right]} \quad ,
\label{runmg}
\end{equation}

\noindent where $m_g \sim 0.5$ GeV \cite{corn82}.

\end{description}

In Table~III we show, for the nucleon case only (results
for the other cases are very similar),
how the three previous different
choices of $\alpha_s$ influence our
results for $a^s_{_B}$ and the decay widths. It is easy to see
that $a^s_{_B}$ is very stable against these
changes (as opposed to the total decay widths),
giving us more confidence on the reliability of our
estimates for $a^s_{_B}$ and on their usefulness.

Note that the non--relativistic $DA$ has not been taken into account,
because in this case there is no ambiguity, in that
the strong coupling constant is consistently fixed to the
effective value $\alpha_s(M_\psi^2)$.

Let us conclude by stressing that we also must take into
account the evolution of the $DA$ with $Q^2$.
It is usually said that this evolution has little effect on the
results, in that its logarithmic behaviour is masked by
the stronger, powerlike behaviour (in $Q^2$) of the elementary
scattering amplitudes. Even if this may seem quite reasonable,
we have explicitly checked for this assumption, particularly
when modified running expressions for $\alpha_s$ have
been adopted. As a matter of fact,
while $a_{_B}$ is almost independent of the $DA$ evolution,
the modifications of the decay widths due to the evolution
are not numerically negligible; however, they are not important
from a qualitative point of view and do not change
the results by more than a factor $\sim$ 2, which is
within the overall uncertainty  we can expect, also
in the more optimistic hypothesis, for absolute quantities.

\section{Electromagnetic contributions \protect\\
to \protect$\bbox{\lowercase{a}_{_B}}$}
\label {aebsec}

Up to now we have neglected the electromagnetic contributions
to $a_{_B}$. However, it has been shown that the corrections due to
e.m. processes can be sizable \cite{cari87,glas82},
and a careful analysis of $a_{_B}$ cannot avoid considering them.
Unfortunately, as it was discussed by Carimalo \cite{cari87},
we are not able,
at present, to give a full treatment of these effects.
The reason is that the time--like e.m. form factors of the
baryons are involved, and there is no calculation
of them including the mass corrections, as it has been done
in this paper for the strong contribution.
Only in the nucleon case, with a non--relativistic $DA$,
this calculation has been performed by Ji {\it et al.} \cite{sill86}.

In our treatment we will follow, expand and update
the analysis made by Carimalo in ref. \cite{cari87}.
After a general discussion of these corrections, we shall consider,
from two different points of view, their estimation
(strongly dependent on the baryon e.m. form factors)
and their contribution to $a_{_B}$, limiting ourselves to
the case of the nucleon. In fact, for the other octet baryons
both experimental and theoretical knowledge of the form
factors is too poor to allow a sensible phenomenological
study.

We first try to give a full theoretical prediction for
$a_{_B}$ in the framework of our model. As we said before,
at present this is possible only for the nucleon and with
a non--relativistic distribution amplitude. However, we
must not forget that non--relativistic $DA$'s
give results far from the experimental ones
for the e.m. hadron form factors (which
are generally underestimated and given,
in some cases, with the wrong sign \cite{brod89}).
So, we must take the non relativistic case
as the only one for which at present
our model can give a full estimate of $a_{_B}$,
together with indications on the weight of e.m. contributions,
waiting for more reliable calculations of mass correction
effects to baryon e.m. form factors in the near future.

Next we shall consider these effects from
a more phenomenological point of view,
trying to estimate their order of magnitude from
the available experimental results on the decay widths
of the $J/\psi$ into $B\bar B$ and $e^-e^+$ pairs, respectively.

The reduced amplitudes defined in Eq.~(\ref{ampli}), including
e.m. contributions, may be written as follows

\begin{equation}
\tilde A_{+\raisebox{-.15ex}{$\scriptstyle\pm$}}
= \tilde A^s_{+\raisebox{-.15ex}{$\scriptstyle\pm$}}
+ \tilde A^{em_1}_{+\raisebox{-.15ex}{$\scriptstyle\pm$}}
+ \tilde A^{em_2}_{+\raisebox{-.15ex}{$\scriptstyle\pm$}}
\quad ,
\label{afull}
\end{equation}

\noindent where both e.m. contributions are of order $(\alpha/\alpha_s)$
with respect to the strong one, and all other higher order
contributions in $(\alpha/\alpha_s)$
are neglected. The term $em_1$ corresponds,
for the hard scattering among the valence quarks,
to Feynman graphs like those of
Fig.~\ref{qcd}, where a gluon is replaced by a photon; there are
three topologically distinct
graphs in this case, because we can replace each gluon by a photon.
The contribution $em_2$ represents the decay of the $J/\psi$ into
a single virtual photon which in turn is directly (through
the time--like e.m. baryon form factors) coupled
to the final baryon pair. Fig.~\ref{qed} shows the Feynman graphs
pertinent to these terms.

Let us consider, first of all, the $em_1$ contribution.
It is not difficult to see that, for the elementary, hard
scattering amplitude, we have

\begin{equation}
T^{em_1}_i = -\frac{4}{5}\frac{\alpha}{\alpha_s}e_i T^s \quad ,
\label{tem1i}
\end{equation}

\noindent where $e_i$ is the electric charge, in units of the
proton charge, of the $i$--th quark, to which the virtual
photon is coupled, and $T^s$ is given by Eq.~(\ref{T}).
As we said, there are three contributions
like this, which sum up to give

\begin{equation}
T^{em_1} = -\frac{4}{5}\frac{\alpha}{\alpha_s}Q_{_B} T^s
= \delta_{_B} T^s \quad .
\label{tem1}
\end{equation}

Here $Q_{_B} = e_1+e_2+e_3$ is the baryon electric charge
(in units of the proton charge); we have also defined the constant
$\delta_{_B}$
($\delta_B = 0$ for neutral baryons and
$\delta_B \sim \pm 10^{-2}$ for $Q = \pm 1$ baryons).
The subsequent steps which lead to the physical
amplitudes $\tilde A$ are the same as for the strong
contribution, so the same relation of
Eq.~(\ref{tem1}) applies also to them:

\begin{equation}
\tilde A^{em_1}_{+\raisebox{-.15ex}{$\scriptstyle\pm$}}
= \delta_{_B} \tilde A^s_{+\raisebox{-.15ex}{$\scriptstyle\pm$}}
\quad .
\label{aem1}
\end{equation}

The case of the $em_2$ contribution is more subtle.
First of all, given that the virtual photon couples
directly to the final baryons, there is not an equivalent
of the hard elementary amplitudes $T$, as in the preceding cases
( or better, this step of the
calculation is hidden in the baryon form factors).
Then we must start from the decay of two free
$c$, $\bar c$ quarks into the final baryon pair, which is
described by means of the $M$ amplitudes we defined,
for the strong contribution,  in Sect. \ref{asbsec}.

It can be seen that

\begin{eqnarray}
\lefteqn{ M^{em_2}_{\lambda_{_B}\lambda_{_{\bar B}};
\lambda_c\lambda_{\bar c}}(\theta_{_B}=\varphi_{_B}=0) = }
\nonumber \\
& &\ \frac{8}{\sqrt{3}}\pi\alpha\frac{1}{M_\psi}
\biggl\{ (F^B_1+k_{_B}F^B_2)  \nonumber \\
& & \times \left[ 2\delta_{\lambda_c\lambda_{_B}}
\delta_{\lambda_{c},-\lambda_{\bar c}}
\delta_{\lambda_{_B},-\lambda_{_{\bar B}}} +
2\epsilon_{_B}
\delta_{\lambda_{c},\lambda_{\bar c}}
\delta_{\lambda_{_B}\lambda_{_{\bar B}}} \right]
\nonumber \\
& & + \frac{k_{_B}}{2\epsilon_{_B}}
F^B_2(1-4\epsilon^2_{_B})
\delta_{\lambda_{c},\lambda_{\bar c}}
\delta_{\lambda_{_B}\lambda_{_{\bar B}}}
\biggr\} \quad ,
\label{mem2}
\end{eqnarray}

\noindent where $k_{_B}$ is the baryon anomalous magnetic moment
and $F^B_{1,2}$ are the well known Dirac and Pauli baryon form factors
respectively,
at $q^2 = M_\psi^2$.

Then, from Eqs.~(\ref{atfroma1}),(\ref{atfroma2}),(\ref{A1}),
(\ref{A2}) we obtain

\begin{equation}
\tilde A^{em_2}_{+-} = \frac{2^{9/2}}{\sqrt{3}}\pi^2
\alpha|R_s(0)|\frac{1}{M_\psi} G^B_M \quad ,
\label{atem2pm}
\end{equation}

\begin{equation}
\tilde A^{em_2}_{++} = \frac{2^5}{\sqrt{3}}\pi^2
\alpha|R_s(0)|\frac{1}{M_\psi}\epsilon_{_B} G^B_E \quad ,
\label{atem2pp}
\end{equation}

\noindent where $G^B_M = F_1^B+k_B F_2^B$ and
$G^B_E = F_1^B+ (q^2/4m_{_B}^2)k_B F_2^B$
are the Sachs baryon magnetic and electric
form factors respectively, again at $q^2=M_\psi^2$.

We are now equipped with a formalism for calculating $a_{_B}$
including the leading electromagnetic corrections.
Unfortunately a model consistent
evaluation requires the  analytical expressions for the baryon
form factors including mass corrections, which at present are
available only in the particular case of the non--relativistic
$DA$.

{}From the paper of Ji {\it et al.} \cite{sill86} we find

\begin{equation}
G^p_M(s) = 54F_N^2\pi^2\alpha_s^2\frac{1}{s^2}
(-3-78\epsilon^2_{p}+136\epsilon^4_{p}) \quad,
\label{gmp}
\end{equation}

\begin{equation}
G^p_E(s) = 54F_N^2\pi^2\alpha_s^2\frac{1}{s^2}
(-11-24\epsilon^2_{p}+48\epsilon^4_{p}) \quad ,
\label{gep}
\end{equation}

\begin{equation}
G^n_M(s) = 18F_N^2\pi^2\alpha_s^2\frac{1}{s^2}
(9+140\epsilon^2_{n}-272\epsilon^4_{n}) \quad ,
\label{gmn}
\end{equation}

\begin{equation}
G^n_E(s) = 108F_N^2\pi^2\alpha_s^2\frac{1}{s^2}
(7-10\epsilon^2_{n}) \quad .
\label{gen}
\end{equation}

It is easy to check that these results reproduce the
corresponding massless results,when $\epsilon_{p,n} \to 0$
(see, {\it e.g.}, ref.~\cite{cer84b} and references
therein).
However, we point out that
mass corrections seem to be greater than expected,
in that in some cases the higher order terms in
$\epsilon_{p,n}^2$
are larger than the leading ones. We never encountered
such a situation in our own calculations of mass corrections
effects.

By insertion of these results (at $s = M^2_\psi$) and Eq.~(\ref{aem1}) into
Eq.~(\ref{afull}) we can estimate the value of $a_{_B}$
and its variations with respect to the calculation
made only with the strong contribution.
The results are reported in Table~IV,
together with the values of the decay widths $\Gamma(J/\psi \to B\bar B)$,
both for the
proton and the neutron. We see that percentual variations of $a_{_B}$
with respect to the pure strong contribution case are
almost negligible:
the effects of e.m. contributions seem to be
much smaller than those due to the choice of different $DA$'s,
see Table~I.

Let us consider now what can be said on the e.m.
contributions from a phenomenological analysis,
based upon the experimental information on some ingredients
of the model.

{}From Eqs.~(\ref{ab2}),(\ref{afull}),(\ref{aem1}),
(\ref{atem2pm}),(\ref{atem2pp}) we can see that:

\begin{equation}
a_{_B} = \frac{1-\rho+a^s_{_B}(1+\rho)}{1+\rho+a^s_{_B}(1-\rho)} \quad ,
\label{acorr}
\end{equation}

\noindent where we have defined

\begin{equation}
\rho = \left(\frac{1+zx_{+-}}{1+x_{+-}} \right)^2 \quad ,
\label{rho}
\end{equation}

\begin{equation}
z = 2\left( \frac{1+a^s_{_B}}{1-a^s_{_B}}\right)^{1/2}
\epsilon_{_B}\frac{G^B_E}{G^B_M}
\label{z}
\end{equation}

\noindent and

\begin{equation}
x_{+-} = \frac{1}{1+\delta_{_B}}
        \frac{\tilde A^{em_2}_{+-}}{\tilde A^s_{+-}} \quad .
\label{x}
\end{equation}

We see then that in order to estimate $a_{_B}$ we need to
know $a^s_{_B}$ and the ratios $G^B_E/G^B_M$ and $x_{+-}$.
$a^s_{_B}$ has been evaluated in Sections \ref{asbsec},\ref{asbris},
while for the two ratios we must resort to experimental
information.

Experimental estimates of
$x_{+-}$ can be obtained as follows:
from Eqs.~(\ref{gtot}),(\ref{afull}),(\ref{atem2pm}),
(\ref{atem2pp}) we can write

\begin{eqnarray}
\Gamma_{_{B\bar B}} & =  & \Gamma_{ee}(1-4\epsilon^2_{_B})^{1/2}
|G^B_M|^2 \left\{ \left(1+\frac{1}{x_{+-}}\right)^2 \right. \nonumber \\
& + & \left. 2\epsilon^2_{_B}\left|\frac{G^B_E}{G^B_M}\right|^2
\left(1+\frac{1}{zx_{+-}}\right)^2 \right\} \quad ,
\label{perx}
\end{eqnarray}

\noindent where $\Gamma_{_{B\bar B}} = \Gamma(J/\psi \to B\bar B)$,
$\Gamma_{ee} = \Gamma(J/\psi \to e^-e^+)$, and all other
quantities have been previously defined.
{}From Eq.~(\ref{perx}) we can evaluate $x_{+-}$,
once $|G^B_M|$  and the ratio $G^B_E/G^B_M$
have been given. We take the experimental data on
$\Gamma_{_{B\bar B}}$, $\Gamma_{ee}$ from ref.~\cite{pdg92}.

Let us first consider the proton case.
Very recently the time-like proton magnetic form factor
at $q^2 \simeq 10$ GeV$^2$ has been measured \cite{e76093},
$|G_M^p(M^2_\psi)| = 0.026 \pm 0.002$
(experimental results
on $G^p_M$ were not available at the time ref. \cite{cari87}
was published. By analytic continuation, it was assumed that
to leading order in $\alpha_s$, $G^p_M(q^2=M_\psi^2)
\simeq G^p_M(q^2=-M_\psi^2) \simeq 0.012$.
This does not seem to be the case \cite{e76093}).

So, for the proton the only quantity we
require which is not experimentally
measured at present is the ratio $|G^B_E/G^B_M|$.
One possibility is to assume approximate validity of the
empirical relation $G^p_E/G^p_M \simeq 1/\mu_p$.
However, there is nothing assuring that this relation
can be verified in the kinematical regime of interest here. For such
reason, we will subsequently take this ratio as a free parameter,
allowing for reasonable variations of its value
around the expected, empirical one, in order to
study the dependence of $a_{_B}$  from this uncertainty.

Given that our results depend on the values of $G^B_M$
and $|G^B_E/G^B_M|$ separately, let us clarify how measurements
have been taken from ref.~\cite{e76093}.
It is seen from ref.~\cite{e76093}
that the measured value of $|G_M^p|$ depends on the value of
the ratio $|G^p_E/G^p_M|$. This is so because there is not enough
statistics in order to derive separate expressions of
$G^p_M$ and $G^p_E$.
So, in our phenomenological analysis we proceed as follows:
once the ratio $G^p_E/G^p_M$ has been fixed to some value,
we can estimate $G^p_M$
from the experimental
data (see Eq.~(4) and Table~I of ref.~\cite{e76093})
and using the same fitting procedure of ref.~\cite{e76093}
to get the form factors
at $s=M_\psi^2$.
The first two columns of Table~V
give corrected values of $a_p$ and percentual variation with respect
to the pure strong contribution, when use is made of the
empirical relation $G^p_E/G^p_M \simeq 1/\mu_p$.
We can see that e.m. corrections can be relevant, and we
probably cannot neglect them in a complete evaluation of $a_{_B}$.
We stress however that these estimates depend on the
theoretical value of the strong contribution
(see Eqs.~(\ref{acorr}-\ref{z})), as it can
be seen from the spread of $\Delta a_p$ over the various $DA$'s
considered. However, since we cover a large range of possibilities
for the strong contribution, we think the order of magnitude
of e.m. corrections can be safely estimated to be
of 5-10 \%, for the proton.
These estimates must be taken as upper limits.

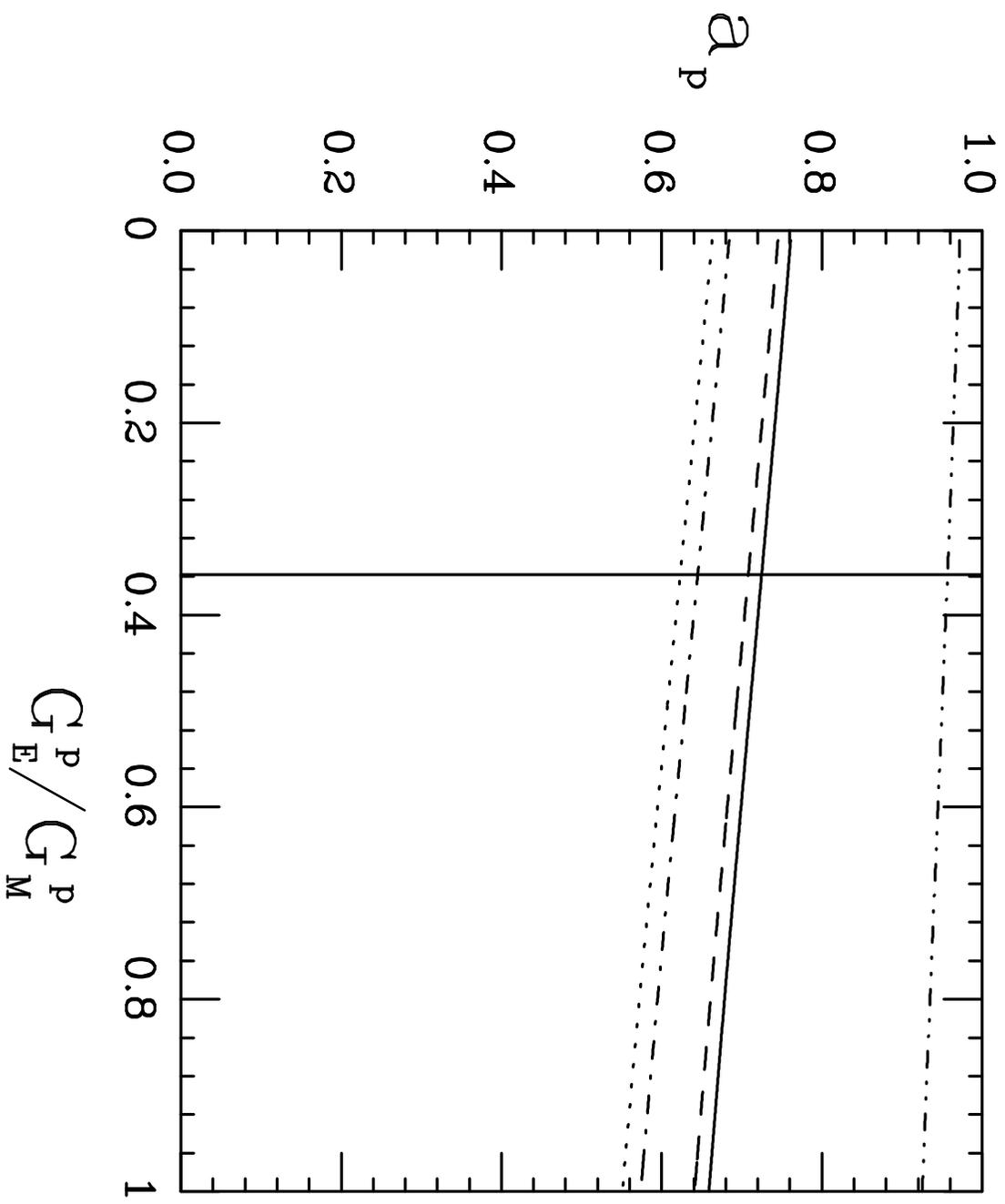
\begin{figure}
\caption{ The dependence of the full (QCD+QED) angular
distribution parameter $a_{_B}$ (see Eq.~(\protect\ref{ab2})) from the
poorly known $G_E/G_M$ ratio, in the proton case.
The vertical, solid line corresponds to the well known
empirical behavior $G_E^p/G_M^p \simeq 1/\mu_p$.
Different lines correspond to the various distribution
amplitudes considered for the strong contribution (see text):
$NR$ (solid); $ASY$ (dashed); $COZ$ (dotted); $KS$ (dot-dashed);
$GS$ (double dot-dashed). In this plot, the $CZ$ and $HET$ $DA$
results are almost indistinguishable from the $COZ$ and
$NR$ ones, respectively. }
\label{two}
\end{figure}

Fig.~\ref{two} shows the variation of $a_p$
as a function of the ratio
$G^p_E/G^p_M$ and for some indicative $DA$'s. The central
value correspond to the possible expected behavior,
$G^p_E/G^p_M \simeq 1/\mu_p$.

We see from Fig.~\ref{two} that our previous conclusions are
not greatly modified; only for big variations of
the behavior of the $G^p_E/G^p_M$ ratio from the
empirical one, we can have relevant differences.
Our conclusion is that the full range of variation
of $a_p$ with respect to $a^s_p$ is of the
order of 0-15 \%.

In the case of the neutron the experimental
information is also poorer than for the proton.
We limit ourselves to extrapolate well known phenomenological
behavior of the neutron form factors to the kinematic
region $q^2 \simeq M_\psi^2$. From this point of view,
we can at first assume $G^n_E \simeq 0$ and
$G^n_M/\mu_n \simeq G^p_M/\mu_p$, at $q^2 = M_\psi^2$.
The last two columns of Table~V show the corresponding
results for $a_n$ and its percentual variation with
respect to the pure strong contribution $a^s_n$.
We see that this time e.m. corrections induce a
decrease of $a_n$, because $G^n_M$ is negative.
As for the absolute variation,
this is slightly larger than in the proton case,
and in the range of 9-15 \%.
Even if we do not report here the results,
we have analyzed  also for the neutron how e.m. corrections change
when $G^n_M$ and $G^n_E/G^n_M$ are allowed to vary around the empirical
values given above. We find again a situation similar
to that of the proton, possibly with a greater spread of
variation.

\section{Conclusions}
\label {conc}

The study of exclusive processes involving hadrons at high
energy scales, in the framework of
perturbative QCD models, has made big improvements in the
last years. However, several problems remain open and
questionable, given that almost all the experimental
information concerning exclusive processes is at
intermediate energy scales. In such situation, several
higher order corrections may not be negligible and
can contribute significantly, at least for those processes
that are forbidden to lowest order in the PQCD models,
but are experimentally well established.
Unfortunately the implementation of these higher twist
effects is very intricate, and only very recently significant
progress has been made. Among these contributions,
mass corrections effects for the valence constituent quarks
of the light hadrons involved are very promising.
In fact, if from one hand a more rigorous theoretical explanation
of their exact origin is required and auspicable,
on the other hand their implementation is relatively easy
and free of ambiguities: once their contribution is taken
into account there are no free parameters to be fixed,
and everything goes on without any further assumptions.
A number of exclusive processes (in particular charmonium decays)
have been analyzed in this framework in the last years,
and several interesting consequences, experimentally testable
at present or in the near future, have been proposed to check
the validity of this model as compared to the PQCD ones
or to alternative models including higher order effects.
In this paper we have analyzed the effects of mass corrections
on the angular distribution of octet baryon pairs created in
$J/\psi$ decays.
Even if experimental measurements of these angular distributions
have, for the moment, a low statistics (at least in some cases)
we hope they can be improved in the near future.

This was also the subject of an earlier
paper by Carimalo. However Carimalo used, as a first approach,
a simplified, non-relativistic model for the produced baryons.
Over time it has been
shown that the behavior of light hadrons, like the octet baryons,
probably demands more accurate models of the baryon
distribution amplitudes,
like those we have considered here in a more general context.
{}From a theoretical point of view, there are several reasons
why the study of these angular distributions is interesting.
First of all they are governed by a parameter
which is given as a ratio of helicity amplitudes.
As such, this parameter results
to be quite independent of some details of
QCD models presently at our disposal, details which are not fully
understood and can modify sizably the
numerical results (albeit not the qualitative ones).
Secondly, as we have shown, this parameter is sufficiently
sensitive to the precise form of the distribution amplitudes
to allow a discrimination among their main features,
as soon as higher precision measurements will be available.
Considering all the available models for the distribution
amplitudes (mainly based on QCD sum rules calculations)
we have shown indeed that the spread on the $a_{_B}$ values
due to the change of the $DA$ is of the order of 10-20 \%,
in the nucleon case. Little is known about
the $DA$'s of the other octet baryons.
Whenever possible, our calculations show an even
larger dependence of $a_{_B}$ from the exact form of the
$DA$, varying in the range of 20-50 \%.
Therefore a more accurate measurement
of $a_{_B}$ could allow us to discriminate among different
model $DA$'s.

We have also considered in detail the role played by
electromagnetic corrections.
In fact, we cannot evaluate exactly these corrections at present.
However, we give reasonable upper limit estimates of these
contributions in the nucleon case, making use of all the possible
available informations , both theoretical
and experimental.

Theoretical estimates (using a non--relativistic
approximation for the final baryon distribution amplitudes)
suggest that e.m. corrections could be negligible,
both in the proton and neutron case.

On the contrary, we estimate from a phenomenological analysis
that e.m. corrections could be of the order of 5-10 \% for the proton
and of 10-15\% for the neutron. In the proton case
e.m. corrections tend to increase the value of $a_{_B}$,
while for the neutron they produce the opposite
effect. In both cases, these corrections
could be comparable to the modifications
induced in the pure strong contribution $a^s_{_B}$
by different choices of the $DA$.

We can say very little about e.m. corrections for
the other octet baryons, since little is known about them, both
experimentally and theoretically.
So we limit ourselves to report, when possible,
the pure strong contribution.

Of course, only more precise experimental
information on the e.m. baryon form factors could allow
to evaluate the exact contribution of e.m. effects.
Alternatively, the form factors could be evaluated
in the framework of our model (including
mass corrections), and a more consistent,
theoretical result could be obtained. We leave this
as a subject for future work.

\acknowledgments

It is a pleasure to thank M.~Anselmino and A.~Devoto
for many useful discussions and for a critical reading of
the manuscript.


\begin{figure}
\caption{ The two Feynman graphs which describe the leading
(in $\alpha/\alpha_s$) electromagnetic
corrections to the QCD lowest order term (see Fig.~\protect\ref{qcd}):
a) The contribution $em_1$, obtained by substituting
one of the virtual gluons in Fig.~\protect\ref{qcd} with a photon. There are
two other contributions, obtained by replacing each of the other two gluons
with a photon (see Fig.~\protect\ref{qcd} for the notation).
b) The contribution $em_2$, coming from the direct
coupling of a virtual photon produced in the $Q \bar Q$
decay with the final baryons. }
\label{qed}
\end{figure}

\narrowtext

\begin{table}
\caption{The strong contribution to $a_{_B}$ and to the decay
width $\Gamma(J/\psi \to B\bar B)$ for the nucleon.
Results obtained using the different nucleon
$DA$'s considered in the text are compared to experimental data.}
\begin{tabular}{cdd}
$DA$ & $a^{s}_{N}$ &
$\Gamma^{s}_{N\bar{N}} \cdot 10^{7}$(GeV) \\
\tableline
$NR$ & 0.688 & 0.002 \\
$AS$ & 0.667 & 0.026 \\
$CZ$ & 0.561 & 0.587  \\
$COZ$ & 0.565 & 0.826  \\
$KS$ & 0.591 & 1.255 \\
$GS$ & 0.963 & 0.168 \\
$HET$ & 0.689 & 1.671  \\
 & & \\
MK2\tablenote{MARKII Collaboration,
ref.~\protect\cite{mark84}.}
& \multicolumn{1}{c}{0.61 $\pm$ 0.23} & \multicolumn{1}{c}
{1.85 $\pm$ 0.27} \\
DM2\tablenote{DM2 Collaboration,
ref.~\protect\cite{dm287}.}
& \multicolumn{1}{c}{0.62 $\pm$ 0.11} & \multicolumn{1}{c}
{1.63 $\pm$ 0.38} \\
\end{tabular}
\label{table1}
\end{table}

\mediumtext

\begin{table}
\caption{The strong contribution to $a_{_B}$ and to the decay
width $\Gamma(J/\psi \to B\bar B)$ for the $\Sigma^+$,
$\Xi^-$ and $\Lambda$.
Only the non--relativistic and the $COZ$ $DA$'s are available
for these particles.}
\begin{tabular}{cdddddd}
 & \multicolumn{2}{c}{$\Sigma^{+}$\tablenote{Experimental results
are available only for the $J/\psi \to \Sigma^0\bar\Sigma^0$ case.}}
& \multicolumn{2}{c}{$\Xi^{-}$} &
\multicolumn{2}{c}{$\Lambda$} \\
\tableline
$DA$ & $a_{B}^{s}$ &
$\Gamma^{s}_{B\bar{B}} \cdot 10^{7}$ &
$a_{B}^{s}$ &
$\Gamma^{s}_{B\bar{B}} \cdot 10^{7}$ &
$a_{B}^{s}$ &
$\Gamma^{s}_{B\bar{B}} \cdot 10^{7}$ \\
 & & (GeV) & & (GeV) & & (GeV)  \\
\tableline
$NR$ & 0.431
& 0.002 & 0.274
& 0.002
& 0.513 & 0.003 \\
$AS$ & 0.417
& 0.032 & 0.265
& 0.016
& 0.497 & 0.017 \\
$COZ$ & 0.687
& 58.151 & 0.537
& 45.519
& 0.770 & 7.252 \\
 & & & & & & \\
MK2\tablenote{MARKII Collaboration,
ref.~\protect\cite{mark84}.}
&\multicolumn{1}{c}{0.7 $\pm$ 1.1}
&\multicolumn{1}{c}{1.35 $\pm$ 0.35}
&\multicolumn{1}{c}{ $-$0.13 $\pm$ 0.55}
&\multicolumn{1}{c}{ 0.97 $\pm$ 0.25}
&\multicolumn{1}{c}{ 0.72 $\pm$ 0.36}
&\multicolumn{1}{c}{ 1.35 $\pm$ 0.27} \\
DM2\tablenote{DM2 Collaboration,
ref.~\protect\cite{dm287}.}
&\multicolumn{1}{c}{0.22 $\pm$ 0.31}
&\multicolumn{1}{c}{0.91 $\pm$ 0.27}
&\multicolumn{1}{c}{}
&\multicolumn{1}{c}{0.60 $\pm$ 0.15}
&\multicolumn{1}{c}{0.62 $\pm$ 0.22}
&\multicolumn{1}{c}{1.18 $\pm$ 0.26} \\
\end{tabular}
\label{table2}
\end{table}

\widetext

\begin{table}
\caption{ Dependence of the strong contribution to $a_{_B}$
and to $\Gamma(J/\psi \to B\bar B)$
on different behaviors of the strong coupling constant
$\alpha_s$ inside the convolution integral of Eq.~(\protect\ref{M}),
in the proton case.
First column is the same as Tab.~I, with $\alpha_s$ from
Eq.~(\protect\ref{run}) at $Q^2 = M_\psi^2$; in the 2nd and 3rd column use
is made of Eq.~(\protect\ref{runcut}), with $\alpha_0 = 0.3$
and $0.5$, respectively; the 4th column presents the results
when Eq.~(\protect\ref{runmg}) is used.}
\begin{tabular}{cdddddddd}
 & \multicolumn{2}{c}{$\alpha_{s}=0.275$} &
\multicolumn{2}{c}{$\alpha_{s} \leq 0.3$} &
\multicolumn{2}{c}{$\alpha_{s} \leq 0.5$} &
\multicolumn{2}{c}{$\alpha_{s}(m_{g}^{2})$} \\
\tableline
$DA$ & $a_{N}^{s}$ &
$\Gamma^{s}_{N\bar{N}} \cdot 10^{7}$ &
$a_{N}^{s}$ &
$\Gamma^{s}_{N\bar{N}} \cdot 10^{7}$ &
$a_{N}^{s}$ &
$\Gamma^{s}_{N\bar{N}} \cdot 10^{7}$ &
$a_{N}^{s}$ &
$\Gamma^{s}_{N\bar{N}} \cdot 10^{7}$ \\
 & & (GeV) & & (GeV) & & (GeV) &  & (GeV) \\
\tableline
$AS$ & 0.667 & 0.026 & 0.667 & 0.044 & 0.666 & 0.466 &
0.667 & 0.210 \\
$CZ$ & 0.561 & 0.587 & 0.561 & 1.144 & 0.564 & 10.013 &
0.561 & 5.716 \\
$COZ$ & 0.565 & 0.826 & 0.565 & 1.369 & 0.567 & 12.078 &
0.564 & 6.894 \\
$KS$ & 0.591 & 1.255 & 0.591 & 2.076 & 0.592 & 18.343 &
0.591 & 10.549 \\
$GS$ & 0.963 & 0.168 & 0.963 & 0.275 & 0.954 & 1.917 &
0.963 & 1.160 \\
$HET$ & 0.689 & 1.671 & 0.686 & 2.763 & 0.682 & 24.245 &
0.691 & 14.020 \\
\end{tabular}
\label{table3}
\end{table}

\narrowtext

\begin{table}
\caption{Theoretical predictions for $a_{_B}$ and
the decay width $\Gamma_{B\bar B}(J/\psi \to B\bar B)$ , including
e.m. corrections, for the proton and the neutron.
The 2nd column gives the percentual variation of $a_{_B}$
with respect to the pure strong contribution.
Only the non--relativistic $DA$ is considered, and the results of Ji
{\it et al.}
\protect\cite{sill86} for the nucleon e.m. form factors are used. }
\begin{tabular}{cddd}
 & $a_{N}$ & $\Delta a_{N}$(\%) & $\Gamma_{N\bar{N}} \cdot 10^{7}$(GeV) \\
\tableline
$p$ & 0.696 & 1.2 & 0.002 \\
$n$ & 0.677 & $-$1.4 & 0.002 \\
\end{tabular}
\label{table4}
\end{table}

\narrowtext

\begin{table}
\caption{Results for $a_{p,n}$ obtained by adding to the
theoretical values of $a^s_{p,n}$ phenomenological estimates
of the e.m. corrections. All the proposed $DA$'s are considered.
The $G^{p,n}_M(M_\psi^2)$ and $G^{p,n}_E(M_\psi^2)/G^{p,n}_M(M_\psi^2)$
values are fixed by using, when available, experimental data
or by extrapolating them to the kinematical region under
study (see text for details).}
\begin{tabular}{cdddd}
 & \multicolumn{2}{c}{$p$} & \multicolumn{2}{c}{$n$} \\
\tableline
$DA$ & $a_{p}$ &
$\Delta a_{p}$(\%) &
$a_{n}$ &
$\Delta a_{n}$(\%) \\
\tableline
$NR$ & 0.725 & 5.3 & 0.628 & $-$8.9 \\
$AS$ &0.708 & 5.9 & 0.605 & $-$9.7 \\
$CZ$ & 0.620 & 10.1 & 0.483 & $-$14.7 \\
$COZ$ & 0.623 & 9.9 & 0.487 & $-$14.6 \\
$KS$ & 0.645 & 8.8 & 0.517 & $-$13.2 \\
$GS$ & 0.956 & $-$0.7 & 0.954 & $-$0.8 \\
$HET$ & 0.726 & 5.3 & 0.629 & $-$8.9 \\
\end{tabular}
\label{table5}
\end{table}

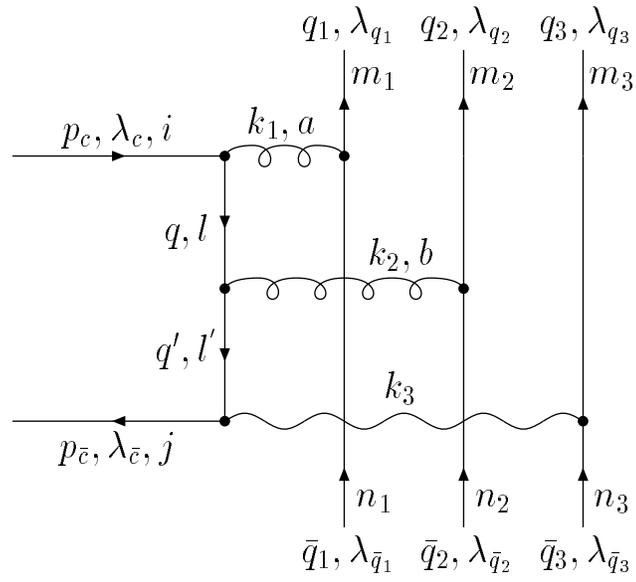
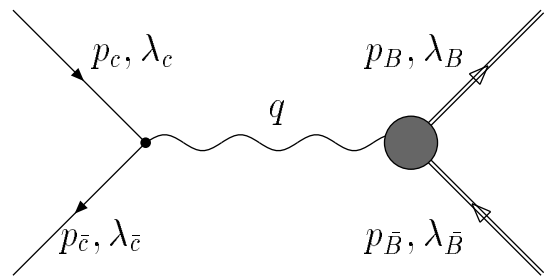
\end{document}